\def\figsize{9.5cm}

\def\rn{}
\def\nn#1 #2{#2. #1}                            % Name with 1 initial
\def\nnn#1 #2 #3{#2. #3. #1}                    % Name with 2 initials
\def\nnnn#1 #2 #3 #4{#2. #3. #4 #1}             % Name with 3 initials
\def\nnnnn#1 #2 #3 #4 #5{#2. #3. #4 #5. #1}     % Name with 4 initials
\def\rf#1;#2;#3;#4;#5 {{\frenchspacing\par\rn#1, #3 {\bf #4}, #5 (#2). \par}}
\def\rrf#1;#2;#3;#4;#5 {{\frenchspacing\rn#1, #3 {\bf #4}, #5 (#2);~}}
\def\rrrf#1;#2;#3;#4;#5 {{\frenchspacing\rn#1, #3 {\bf #4}, #5 (#2).}}
\def\rg#1;#2;#3;#4;#5;#6 {{\frenchspacing\par\rn#1, #3 {\bf #4}, #5 (#2). \par}}
\def\rfbook#1;#2;#3;#4;#5 {{\frenchspacing\par\rn#1, {\it #3} (#5, #4, #2).\par}}
\def\rfprep#1;#2;#3 {{\par\frenchspacing\rn#1, #3 (#2).\par}}
\def\rrfprep#1;#2;#3 {{\frenchspacing\rn#1, #3 (#2);~}}
\def\rrrfprep#1;#2;#3 {{\frenchspacing\rn#1, #3 (#2).}}
\def\rfproc#1;#2;#3;#4;#5;#6 {{\frenchspacing\par\rn#1 #2, in {\it #3}, ed. #4 (#5: #6)\par}}
\def\rfprocp#1;#2;#3;#4;#5;#6;#7 {{\frenchspacing\par\rn#1 #2, in {\it #3}, ed. #4 (#5: #6), p#7\par}}
\def\rg#1;#2;#3;#4;#5;#6 {\par\rn#1 #2, {\it #3}, {\bf #4}, #5 (``#6'') \par}
\def\rf#1;#2;#3;#4;#5 {\par\rn#1, {\it #3}, {\bf #4}, #5 (#2)\par}
\def\rfbook#1;#2;#3;#4;#5 {{\frenchspacing\par\rn#1, {\it #3} (#4: #5, #2)\par}}
\def\rfproc#1;#2;#3;#4;#5;#6 {{\frenchspacing\par\rn#1 #2, in {\it #3}, ed. #4 (#5: #6)\par}}
\def\rfprocp#1;#2;#3;#4;#5;#6;#7 {{\frenchspacing\par\rn#1 #2, in {\it #3}, ed. #4 (#5: #6), p#7\par}}
\def\rfprep#1;#2;#3  {{\par\rn#1, #3, #2\par}}
\def\rfprepp#1;#2;#3 {{\par\rn#1 #2, #3\par}}

\def\beq#1{\begin{equation}\label{#1}}
\def\eeq{\end{equation}}
\def\beqa#1{\begin{eqnarray}\label{#1}}
\def\eeqa{\end{eqnarray}}

\def\spose#1{\hbox to 0pt{#1\hss}}
\def\simlt{\mathrel{\spose{\lower 3pt\hbox{$\mathchar"218$}}
     \raise 2.0pt\hbox{$\mathchar"13C$}}}
\def\simgt{\mathrel{\spose{\lower 3pt\hbox{$\mathchar"218$}}
     \raise 2.0pt\hbox{$\mathchar"13E$}}}
\def\simpropto{\mathrel{\spose{\lower 3pt\hbox{$\mathchar"218$}}
     \raise 2.0pt\hbox{$\propto$}}}

\def\ed{\end{document}}

\def\beq#1{\begin{equation}\label{#1}}
\def\eeq{\end{equation}}
\def\beqa#1{\begin{eqnarray}\label{#1}}
\def\eeqa{\end{eqnarray}}

\documentclass[twocolumn,amsmath,nofootinbib]{revtex4} % For astro-ph
\usepackage{graphicx}% Include figure files
\usepackage{dcolumn}% Align table columns on decimal point
\usepackage{bm}% bold math
\usepackage{url}

\newcommand{\m}{\medbreak}

\newcommand{\no}{\noindent}
\newcommand{\EQ}{\begin{equation}}
\newcommand{\EQA}{\begin{eqnarray}}
\newcommand{\eqa}{\end{eqnarray}}

\newcommand{\AR}{\renewcommand {\arraystretch}{1.5}
\begin{array}{l}}
\newcommand{\bAR}{\renewcommand {\arraystretch}{2}
\begin{array}{l}}
\newcommand{\ARc}{\renewcommand {\arraystretch}{1.5}
\begin{array}{c}}
\newcommand{\bARc}{\renewcommand {\arraystretch}{2}
\begin{array}{c}}
\newcommand{\ar}{\end{array} \renewcommand {\arraystretch}{1}}

\begin{document}

%%%%%%%%%%%%%%%%%%%%%%%%%%%%%%%%%%%%%%%%%%%%%%%%%%%%%%%%%%%%%
%%%%%%%%%%%%%% new (documentclass) %%%%%%%%%%%%%%%%%
\input{epsf.sty}
\begin{titlepage}
\vspace*{.15in}
%\noindent
%PHYSICAL REVIEW LETTERS
%\vskip .20in
%\noindent

%%%%%
%%%%%
%\today
%\vspace{0.2in}
%\begin{flushright}
%%draft1 AE+CT\\
%\end{flushright}
\vspace*{0.5cm}
\begin{center}
{\Large \bf On the determination of the deceleration parameter
from Supernovae data}\\

\m
\vspace*{1.2cm}
{\bf J.-M. Virey$^1$, P. Taxil$^1$, A. Tilquin$^2$, A. Ealet$^2$, C. Tao$^2$ and D. Fouchez$^2$
 }

\vspace*{1.2cm}
$^1$Centre de Physique Theorique$^*$, CNRS-Luminy,
Case 907, F-13288 Marseille Cedex 9, France \\
and Universite de Provence \\ 
\vspace*{0.5cm}
$^2$Centre de Physique des Particules de Marseille$^+$, CNRS-Luminy,
Case 907, F-13288 Marseille Cedex 9, France

\vspace*{2.5cm}
{\bf Abstract} \\
\end{center}

Supernovae searches have shown that a simple matter-dominated and decelerating
universe should be ruled out. However a determination of the present deceleration
parameter $q_0$ through a simple kinematical description is not exempt
of possible drawbacks. We show that, with a time 
dependent equation of state for the dark energy, a bias is present
for $q_0$ : models which are very far
from the so-called Concordance Model can be accommodated by the data
and a simple kinematical analysis
can lead to wrong conclusions. We present a quantitative treatment of this bias
and we present our conclusions when a possible dynamical dark energy is taken into account.

\vfill
\begin{flushleft}
PACS Numbers : 98.80.Es, 98.80.Cq\\
Key-Words : cosmological parameters - supernovae
\m\no
Number of figures : 3\\

\m\no
February 2005\\
CPT-2005/P.007\\
%CPPM-P-2004-02\\
\m\no
anonymous ftp or gopher : cpt.univ-mrs.fr\\ \no
E-mail : virey@cpt.univ-mrs.fr\\ \m

\no ------------------------------------\\ \m
\no $^*$``Centre de Physique Th\'eorique'' is UMR 6207 - ``Unit\'e Mixte
de Recherche'' of CNRS and of the Universities ``de Provence'',
``de la M\'editerran\'ee'' and ``du Sud Toulon-Var''- Laboratory
affiliated to FRUMAM (FR 2291).\\ \m

\no $^+$``Centre de Physique des Particules de Marseille'' is UMR 6550 
of CNRS/IN2P3 and of the University
``de la M\'editerran\'ee''.
%$^{\ast}$Unit\'e  de Recherche yyyy
% \\
\end{flushleft}
\newpage
\end{titlepage}

%%%%%%%%%%%%%%%%%%%%%%%% fin old (documentstyle) %%%%%%%%%%%%%
%%%%%%%%%%%%%%%%%%%%%%%%%%%%%%%%%%%%%%%%%%%%%%%%%%%%%%%%%%%%%%%%%%%%%%%%%%%%
                   %%%%%%%%%%%%%%%%%%%%%%%%%%%%%%%%%%
   
%\pagestyle{myheadings}
%\markright{draft2-AE+CT}  
%%%%%%%%%%%%%%%%%%%%%%%%%%%%%%%%%%%%                         

%\section{Introduction}

\indent
%\m

Observations \cite{newSCP,Riess04} of Type Ia Supernovae (SNe)
allow to probe the expansion history of the universe. 
The measurements of the apparent magnitudes $m(z)$ of SNe
determine the luminosity distance versus redshift 
$d_L(z)$ 
which reads for
a flat ($\Omega_T = 1$) universe :
\begin{eqnarray}
\label{dLq}
d_L(z)&=& c(1+z)\int_0^z {du\over H(u)} \nonumber\\
  &=&
{c(1+z)\over H_0} \int_0^z du \; e^{\, \left[-\int_0^u [1+q(x)]d\ln (1+x) \right]} 
\end{eqnarray}
\noindent
where $d_L(z)$ is written here in terms of the epoch-dependent deceleration parameter
$q(z) \equiv (-\ddot a /a)/H^2(z)$ with $H(z) = {\dot a}/ a$.
$d_L(z)$ is related to the
measured magnitude by $m(z) = M_S + 5 log_{10}(H_0/c \, d_L(z))$, $M_S$ being a normalisation parameter
which combines the absolute SNe magnitude and the Hubble constant.

The present behavior of the expansion is attributed to a new "Dark Energy" (DE)
component with negative pressure : $p_X = w\, \rho_X$, with a possibly
time dependent equation of state (EoS) $w(z)$
 (for a recent review see Ref. \cite{Padma2004}). 
$\Omega_X$ ($\Omega_M$) 
will denote the ratio of the
present DE (Matter) density to the critical density.
In this framework, with $\Omega_T = 1$ and neglecting the radiation component :
\begin{eqnarray}
\label{dLw}
d_L(z)& = &
{c(1+z)\over H_0} \int_0^z du \\
& & \left[ (1+u)^3 \Omega_M +
\Omega_X e^{\, 3\int_0^u [1+w(x)]d\ln (1+x)} \right]^{-1/2} \nonumber 
\end{eqnarray}

The connection with Eq.(\ref{dLq}) is given by :
\begin{equation}
q(z)  =  {1\over 2}+{3\over 2}w(z)\Omega_X(z)
\; ;
%{\rm with}
\; 
\Omega_X(z) = \Omega_X \frac{\rho_X(z) H_0^2}{\rho_X(0) H^2(z)}
\end{equation}

In Ref.\cite{Riess04} Riess et al. have recently presented an
analysis of 156 SNe
including a few at $z > 1.3$ from the Hubble Space Telescope (HST) GOODS ACS Treasury survey.
They use a kinematical description (no dynamical hypothesis) with a simple 
parametrization of $q(z)$ in Eq.(\ref{dLq}), $q(z)=q_0\, +\, q_1\, z$, 
and conclude to the evidence
for present acceleration  $q_0<0$ at $99\%$ C.L. ($q_0 \approx -0.7$)
and for past deceleration $q(z) >0$ beyond $z_t = 0.46 \pm 0.13$. 
Concerning the dynamics in a flat universe,
they conclude on the validity of the 
Cosmic Concordance  version of the $\Lambda$CDM Model that is $\Omega_M \approx 0.3$, $w(z=0) \approx -1$ and no rapid
evolution of the EoS.

Performing the same
kind of analysis, we confirm the numbers and errors obtained with the Gold Sample 
(see Ref. \cite{Riess04}) of SNe 
for the "kinematical fit" :
 $q_0 = - 0.74 \pm 0.18$ and $q_1 = 1.59 \pm 0.63$. In spite of the goodness of the fit and
of the relatively small errors it yields, this simple strategy
leads to some bias which we quantify in the following.

As noticed several times, \cite{Maor,WA,Gerke,LinderB,Virey1} the physical parameters 
$\Omega$'s and $w(z)$
% or $q(z)$) 
are related to the measured quantity $d_L(z)$ through
a multiple integral relation which results in a somewhat
uncertain determination of these parameters, even at their present values
$\Omega_M$ and $w(0)$. 
In Ref.\cite{Virey1} we have analysed quantitatively the bias which occurs
when one tries to extract $w(0)$ neglecting a possible redshift
dependence of $w(z)$. In the same time, other fitted quantities (essentially
$\Omega_M$) are affected, due to strong correlations between $\Omega_M$, $w(0)$
and $dw(z)/dz$. Another pitfall originates from the assumed value (and uncertainty) of the
prior on $\Omega_M$ which is used for the fit 
\cite{Virey2} : an artificial convergence
towards the Concordance Model seems to occur whereas the simulated fiducial model
is very different. 

Concerning the determination of the
deceleration parameter, one encounters
a similar problem :
the extraction of $q_0$
is not independent of the assumed form of the function $q(z)$, a form
which is related (dynamical approach) or not (kinematical approach of Riess et al.)
to the evolution of the DE EoS and to the value of $\Omega_M$.
Indeed, when choosing the kinematical approach,
Riess et al. are in some sense model independent but they must
use a particular {\em description} of the kinematics : e.g. the linear form 
$q(z)=q_0\, +\, q_1\, z$. It is valuable to question this description, 
for instance in the
light of some dynamical models which include an evolution of the DE EoS.

To modelize the evolution of the EoS, 
in the absence of deep physical insight, the choice of
a parametrization is arbitrary. Simplicity imposes a two-parameter
parametrization for $w(z)$, then consistency with the data
from the Cosmic Microwave background (CMB) imposes $w(z) \leq 0$ at high
redshift. Therefore, the simple linear parametrization :
\begin{equation}
\label{wlinear}
w(z) = w_0 + w_1 \, z
\end{equation}
we have used previously \cite{Virey1,Virey2} (with many other authors) is too badly behaved
at high $z$. We prefer to switch to the
parametrization
advocated in \cite{Polarskiparam,LinderA} and which is now widely used in the
literature \cite{WangTegmark,Choudhury,Rapetti,SeljakA,UIS}. 
\begin{equation}
\label{wLinder}
w(z) = w_0 + w_a z/(1+z)
\end{equation}
\noindent

Then from Eq.(\ref{wLinder}),
one gets : 
\begin{equation}
\Omega_X(z) =  \Omega_X \, {H_0^2\over H^2(z)} \, (1+z)^{3(1+ w_0 + w_a)} e^{-3w_a z/(1+z)}
\end{equation}

\noindent and $q(z)$ reads :
\begin{eqnarray}
\label{qzm}
q(z) & = &  {1\over 2} +{3\over 2}(w_0 + w_a {z\over 1+z})\\
& & \times \left[ {1+{\Omega_M\over \Omega_X}
(1+z)^{-3(w_0 + w_a)} e^{3w_a z/(1+z)} } \right]^{-1} \nonumber 
\end{eqnarray}

The behavior of $q(z)$ is plotted in Fig.1 for various models listed
in Table I : it is obvious that
$q(z)$ is in general far from a linear shape.
Even in the $\Lambda$CDM model, $q(z)$ is not exaclty linear
in $z$ as can be seen from Eq.(\ref{qzm}) and Fig.1.

\def\figsize{6.cm}
\begin{figure} 
%\vskip-1.cm
\centerline{\epsfxsize=\figsize\epsffile{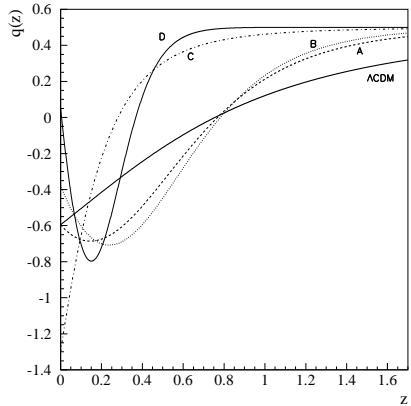}}
%\vskip-1.cm
\caption[1]{\label{fig:fig1}\footnotesize%
$q(z)$ from Eq.(\ref{qzm}) with 
$\Omega_T = 1$ and some particular
values of the parameters ($\Omega_M, w_0, w_a$) given in Table I.
}
\vskip-.4cm
\end{figure}

\begin{table}
%\begin{table}[htbp]
\caption{\protect\footnotesize Model examples 
with fiducial values ($\Omega_M,w_0,w_a$).
The fiducial $q_0(\Omega_M,w_0$) is calculated from Eq.(\ref{qzm}).  
The fitted $q_0$ are obtained with  a linear 3-fit $(M_S,q_0,q_1)$ of the simulated data.
These values are well in agreement with the one extracted from  the Gold Sample
($q_0^d=-0.74\pm 0.18 $). For consistency, Models A,B,C,D have been chosen to give
a good dynamical 4-fit ($M_S,\Omega_M,w_0,w_a$) of the Gold Sample.
}
%\vspace*{0.2cm}
\label{tab:riess}
\begin{tabular}{ |c|c|c|c|c|c|}
\hline 
 model &  $\Omega_M$ & $w_0$  &  $w_a$ & $q_0(\Omega_M,w_0$) & fitted $q_0$   \\
\hline \hline
$\Lambda$CDM & $0.27 $  & $ -1 $  & 0 & $ -0.6$ & $-0.57 \pm 0.17$   \\
\hline
A &  $0.27 $  & $ -1 $  & $-2 $  & $-0.6$ & $- 0.79 \pm 0.17$ \\
B &  $0.27 $ & $-0.8$  & $-3.2 $ & $-0.38$ &  $-0.74 \pm 0.17$   \\
C & $0.50 $ & $-2.4$  & $1.4$  & $-1.3$ & $ -0.75 \pm 0.18$  \\
D & $0.50 $ & $-0.6 $ & $ -15 $  & $0.05$ & $-0.75 \pm 0.18$  \\
\hline \hline
\end{tabular}
%\end{center}
\vspace*{-0.5cm}
\end{table}

To illustrate the consequence of the non-linear form of 
$q(z)$ on the kinematical fit, we simulate SNe data samples
corresponding to the same statistical power as the true data sample by
fixing the fiducial($^F$) values for the parameters $M_S^F, \Omega_M^F,
w_0^F$ and $w_a^F$.
In the following, we consider a flat cosmology and we fix
 $M_S^F = -3.6$. Then we perform a three-parameter ($M_S, q_0, q_1$) kinematical fit of the simulated data. The results for the models we use are presented in Table I. 

First, the $\Lambda$CDM Concordance model ($w_0^F = -1$,  $w_a^F = 0$) 
with the $\Omega_M$ value (0.27) of
\cite{Riess04},
yields a fitted value: $q_0 = - 0.57 \pm 0.17$ which reproduces well the
fiducial value in the limit of the errors. 
Note however that the value $q_0^{data} = - 0.74 \pm 0.18$ obtained from the
real data is 1$\sigma$ away from the actual
$\Lambda$CDM value.

With Model A, where a time variation of the EoS
 is allowed ($w_a^F \neq 0$), 
we illustrate the fact that even if $q_0^F$ from Eq. (\ref{qzm}) is independent of
$w_a^F$, its fitted value is not. 

With model B (this model could correspond to one of the
k-essence models \cite{kessence}), the fitted $q_0$ value is now 2$\sigma$ away from
the fiducial ($q_0^F = -0.38$) . 

In model C, we have changed
the fiducial $\Omega_M^F$ value to 0.5. We have $w_0^F < -1$ 
(as in phantom models \cite{Caldwell}, see e.g. \cite{Carroll1} 
for a list of references) and a positive $w_a^F$. 
The bias is then very large since
the fitted $q_0$ is more than 3$\sigma$  away from the fiducial value.

One can even get 
a truly slowly decelerating model with model D : $q_0^F = 0.05$, still with 
$\Omega_M^F = 0.5$, $w_0^F = -0.6$ 
and a very rapid and recent evolution of $w(z)$ : $w_a^F = -15$.
Of course the behavior of Model D seems somewhat artificial and the result
of the fit can be seen from the non-monotonicity of $q(z)$ in this model
(see Fig.1), which leads to compensations in the integral of Eq.(\ref{dLq}).
However, let
us stress again that Model D as well as  Models A,B,C fit perfectly the present SNe data,
yielding in particular $q_0 \approx q_0^{data}$.

More quantitatively, it is interesting to pin down some regions in the parameter space where 
the kinematical fit gets in trouble. 
Following Ref.\cite{Virey1}, we fix $\Omega_M^F$ and $M_S^F$ and 
we consider the plane of the fiducial ($w_0^F, w_a^F$). 
In this plane, we define the Biased Zone (BZ) for $q_0$ as the zone
where the kinematical fit converges perfectly but where the 
difference between the fitted value and the fiducial value $q_0^F$ is larger
than the statistical error $\sigma (q_0)$. The validity zone (VZ) is the complementary region. 
It must be stressed
that the BZ is undetectable with real data. We give the two zones in Fig. 2
for $\Omega_M^F = 0.27$.
The $\Lambda$CDM falls in the VZ as expected, and the models A,B,C,D
belong to the BZ.
Within the VZ, the zone where the linear approximation $q(z) = q_0 + q_1z$
of Eq.(\ref{qzm}) is acceptable is quite small. In fact the rest of the VZ 
is due to accidental cancellations when Eq.(\ref{qzm}) is injected 
in Eq.(\ref{dLq}).
The lower-right corner of the BZ corresponds to models
such that $w_a \lesssim -3 w_0^2 \Omega_M$ ; they display a non-monotonic
behavior for $q(z)$ which explains the bias.
In addition, even with $\Omega_M^F = 0.27$,
there exists a region in this zone where the conclusion on the sign of $q_0$
could be misleading, i.e. a zone where $q_0 + \sigma (q_0) \leq 0$
whereas $q_0^F \geq 0$. This zone corresponds to $-0.5 < w_0^F < 0$ with large
and negative $w_a^F$. 

\def\figsize{6.cm}
\begin{figure} 
%\vskip-1.cm
\centerline{\epsfxsize=\figsize\epsffile{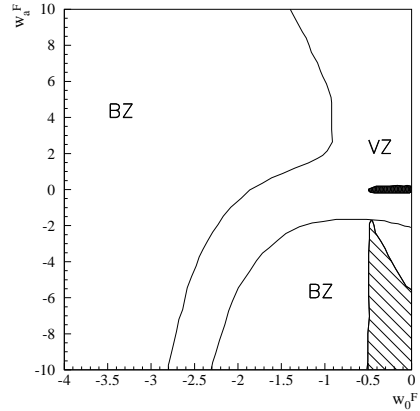}}
%\vskip-1.cm
\caption[1]{\label{fig:fig2}\footnotesize%
Biaised Zone (BZ) and Validity Zone (VZ) for the
deceleration parameter $q_0$, for the kinematical fit ($M_S, q_0, q_1$) of the
simulated data in the
fiducial plane ($w_0^F$,$w_a^F$) with $\Omega_M^F=0.27, M_S^F = -3.6$. 
Small dark zone : where the linear approximation $q(z) = q_0 + q_1 z$ is acceptable ;
hatched zone : where the conclusion on the sign of $q_0$ could be misleading.
}
\vskip-.4cm
\end{figure}

Therefore, it appears that the linear expression of $q(z)$
at first order  : $q(z) = q_0 + q_1 z$, cannot 
be satisfactorily used
over a large range of EoS parameters. 
Going to a second order parametrization will not improve the issue, as it will only enlarge the number of parameters, will degrade the errors, and will not solve the bias problem. 
\\

Alternatively, one can deduce a $q_0$ value from  real data using Eq.(\ref{qzm})
 (with $z=0$) where we consider the 
dynamical description Eqs.(\ref{dLw})and (\ref{wLinder}), by performing a 4-parameter fit 
($M_S,\Omega_M, w_0,w_a$) of the Gold Sample of \cite{Riess04}. 
We have calculated asymetric standard deviations 
(deviation $\delta \chi^2 =1$ for a given parameter),
marginalizing over the other parameters by minimisation. 
For $q_0$, which is a derived parameter, we use a Monte-Carlo technique : generating 1000 simulated experiments around the data best fit, we estimate the width of the $q_0$ distribution. 

\begin{table}
\caption{\protect\footnotesize 
 Results of a "dynamical" 4-fit with $w(z) = w_0 + w_a z/(1+z)$ using the Gold data 
from \cite{Riess04}.
}
\m
%\vspace*{0.2cm}
\label{tab:dyn}
\begin{tabular}{ |c|c|c|c|c|}
\hline 
$\Omega_M$ prior  &   $\Omega_M$ & $w_0$  &  $w_a$ & $q_0$ \\
\hline \hline
$0.27 \pm 0.04$ & $0.27^{+0.04}_{-0.04}$  & $-1.49^{+0.31}_{-0.34}$  
  & $3.20^{+1.60}_{-1.60}$    & $-1.12^{+0.33}_{-0.36}$  \\
& & & &
\\
$0.27 \pm 0.20$ & $0.33 ^{+0.15}_{-0.21}$  
 & $- 1.70^{+0.60}_{-0.60}$    & $3.50^{+2.20}_{-4.30}$  & $-1.21^{+0.49}_{-0.45}$\\
\hline
\end{tabular}
\vspace*{-0.4cm}
\end{table}

The extracted values are displayed in Table II.
In this Table, we have used  a strong prior on $\Omega_M$ ($0.27 \pm 0.04$)
and also a more reasonable weaker uncertainty of $\pm 0.2$, since
adopting strong priors is recognized
to be dangerous (see \cite{Virey2,UIS}).
Comparing with our previous fits,
we see that changing the $w(z)$ parametrization from Eq.(\ref{wlinear})
to Eq.(\ref{wLinder})
has not changed 
the behaviors discussed in \cite{Virey2}. 
In particular choosing a strong prior around $\Omega_M = 0.3$ pushes $w_0$ towards -1 whereas, 
with a weaker  prior 
the central value of $w_0$ is clearly lower than -1, 
a fact which has been noticed by many authors 
(see e.g. \cite{Choudhury,UIS,DicusRepko,Hannestad,Corasaniti}). 

The estimated $q_0$ values confirm  that the Universe is presently in an acceleration phase, at 
$95 \%\ C.L$. We note that the extracted value for a weak prior is  $2\sigma$ away from the 
$\Lambda CDM$ value $q_0 = -0.6$. 
This is a direct consequence of the extracted value of the point ($w_0, w_a$), which is  
$2\sigma$ away from $(-1,0)$. 
With the strong prior, the errors on $q_0$, as on other parameters, are systematically smaller when $\Omega_M$ is close
to the Concordant Model value ($\approx 0.3$). 
This conclusion would be different if we use a larger central value for $\Omega_M$ : 
in this case, the
 $q_0$ value would be even more negative, but with such a large error that 
it is not possible to conclude. 

Finally, if no prior is applied on $\Omega_M$, one 
can conclude with some caution that $q_0 < 0$ at 80\% C.L. 
\\

We can also address the problem of the determination of
the value of the transition redshift $z_t$
between deceleration and acceleration.
Since the kinematical fit yielding the  $q_0$ and $q_1$ values 
found in Ref.\cite{Riess04} is biased, 
one cannot be too confident with the advocated value $z_t = 0.46 \pm 0.13$. 

Fig. 3 displays the function $q(z)=q_{bf}(z)$ computed from the best fit ($bf$) values of the parameters entering  Eq.(\ref{qzm}). The 1$\sigma$ and 2$\sigma$ intervals are computed
by taking the probability density of simulated experiments around the $q_{bf}(z)$ value in each redshift bin.  We still use a weak prior $\Omega_M =0.27 \pm 0.2$. 
The extracted $z_t$ value corresponds to the point where 
$q_{bf}(z_t) = 0$ and $\frac{dq_{bf}}{dz}(z = z_t) >0$. 
We find   
 $z_t = 0.34^{+0.12}_{-0.06}$, where the errors are 
 evaluated from the $z_t$ probability density. 
With the dynamical $q(z)$, 
our errors are small because
the transition is  steeper
than in the linear kinematical case ($q(z) = q_0 + q_1 z$),
as can be seen from Fig. 3. 

Our central value is far
from the "theoretical" $z_t$ of the Concordance Model : 
$ z_t = [{2 \Omega^0_{\Lambda} \over \Omega_M^0}]^{1/3} - 1 = 0.76 
\, ({\rm for}\, \Omega_M = 0.27)$.
Moreover, we have observed that, contrary to the extraction of $q_0$,
the extraction of $z_t$ is very much dependent upon the chosen parametrization
for $w(z)$, a fact mentioned by various authors \cite{Basset,DicusRepko}. 
Then, one can stay prudent and
state that, with present data,  the extracted $z_t$ value is certainly much less robust than
the extracted $q_0$ value, even in a dynamical approach. 
We should also point out that no systematical effects are taken into account in these evaluations, 
in particular a normalisation effect between low and high redshift supernovae would affect directly the  determination of $z_t$.

\def\figsize{6.cm}
\begin{figure} 
%\vskip-1.cm
\centerline{\epsfxsize=\figsize\epsffile{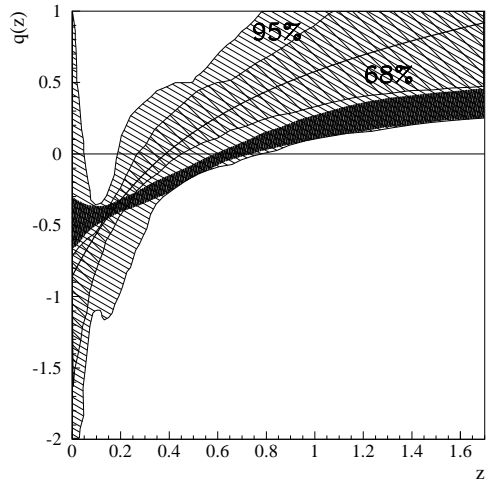}}
%\vskip-1.cm
\caption[1]{\label{fig:fig3}\footnotesize%
best fit $q_{bf}(z)$ function computed with  Eq.(\ref{qzm}) 
using the best fit values of the cosmological parameters 
from the current data (heavy plain curve). The uncertainty contours are constructed from the probability density function around the best fit at a given redshift from one thousand simulated experiments. The dark stripe corresponds 
to the $1\sigma$ error on the fitted function from  a simulation around the 
$\Lambda$CDM concordant model fiducial values, with the SNAP statistics.

}
\vskip-.4cm
\end{figure}

It is interesting to evaluate what should be the best strategy to extract
$q_0$ and $z_t$ in the future.
Using a biased method, as the linear
kinematical one, will be worse when the statistical
errors will decrease. A dynamical, although model-dependent 
approach, remains a good way to estimate
the behavior of the acceleration. 
In Fig. 3,  we show the results 
for a statistical sample expected from the space-based
SNAP mission \cite{LinderSNAP} where
the systematical uncertainties  should be controlled at 
the same level of precision. Thanks to the rich sample of 2000 SNe and also to
the large range of redshift ($0.2 \lesssim z \lesssim 1.7$) 
 a very precise determination of  
the transition region should be allowed.
For instance, with the $\Lambda CDM$ fiducial values ($w_0=-1, w_a = 0, \Omega_M = 0.27$), 
keeping a
weak prior on $\Omega_M$, 
one gets $z_t  = 0.67 ^{+0.08}_{-0.06}$ and $q_0 = -0.55 ^{+0.26}_{-0.13}$. 

We estimate that the correct procedure 
for the future is to avoid using priors on $\Omega_M$. We prefer instead 
the combination of supernovae data with other probes such as the
CMB, the Weak Gravitational Lensing 
and the galaxy power spectrum. 
This kind of method will not introduce external biases and some recent papers go in that
direction (see e.g. \cite{UIS,Hannestad,Corasaniti,Ishak}).

\begin{center}
{\bf Acknowledgments :} 
\end{center}

We thank the members of the Laboratoire d'Astrophysique de Marseille and
Alain Mazure in particular for fruitful discussions.
%$^*$``Centre de Physique Th\'eorique'' is UMR 6207 - ``Unit\'e Mixte
%de Recherche'' of CNRS and of the Universities ``de Provence'',
%``de la M\'editerran\'ee'' and ``du Sud Toulon-Var''- Laboratory
%affiliated to FRUMAM (FR 2291).
%$^+$``Centre de Physique des Particules de Marseille'' is UMR 6550 
%of CNRS/IN2P3 and of the University
%``de la M\'editerran\'ee''.


\vspace*{-0.5truecm}
\begin{thebibliography}{99}


\bibitem{newSCP} R.A. Knop et al.  Astrophys.J. {\bf 598}, 102 (2003)
\bibitem{Riess04} A.G. Riess et al., Astrophys.J. {\bf 607}, 665 (2004)
\bibitem{Padma2004} T. Padmanabhan, Current Science, in press, astro-ph/0411044
\bibitem{Maor} I. Maor, R. Brustein and P.J. Steinhardt, Phys.Rev.Lett.
{\bf 86}, 6 (2001) ; I. Maor et al., Phys.Rev. {\bf D65}, 123003 (2002)
\bibitem{WA} J. Weller and A. Albrecht, Phys.Rev. {\bf D65}, 103512 (2002)
\bibitem{Gerke} B. F. Gerke and G. Efstathiou, MNRAS {\bf 335}, 33 (2002)
\bibitem{LinderB} E. Linder, astro-ph/0406189
\bibitem{Virey1} J.-M. Virey et al.,  Phys.Rev. {\bf D70}, 043514 (2004)
\bibitem{Virey2} J.-M. Virey et al.,  Phys.Rev. {\bf D70}, 121301(R) (2004)
\bibitem{Polarskiparam} M. Chevallier and D. Polarski, Int.J.Mod.Phys. {\bf D10}, 213 (2001)
\bibitem{LinderA} E. V. Linder, Phys. Rev. Lett. {\bf 90}, 091301 (2003)
\bibitem{WangTegmark} Y. Wang and M. Tegmark, Phys.Rev.Lett.
{\bf 92}, 241302 (2004)
\bibitem{SeljakA} U. Seljak et al., astro-ph/0407372
\bibitem{Choudhury} T.R. Choudurhy and T. Padmanabhan, Astron.Astrophys. {\bf 429} 807 (2005)
\bibitem{Rapetti} D. Rapetti, S.W. Allen and J. Weller, astro-ph/0409574 
\bibitem{UIS} A. Upadhye, M. Ishak, P.J. Steinhardt, astro-ph/0411803
\bibitem{kessence} C. Armendariz-Picon et al., Phys.Rev.Lett.{\bf 85}, 4438 (2000); 
Phys.Rev.{\bf D63}, (2001) 103510
\bibitem{Caldwell} R. Caldwell, Phys.Lett.{\bf B545}, 23 (2002) 
\bibitem{Carroll1} S.M. Carrol, A. De Felice and M. Trodden, astro-ph/0408081
\bibitem{DicusRepko} D.A. Dicus and W. W Repko, Phys. Rev. {\bf D70}, 083527 (2004)
\bibitem{Hannestad} S. Hannestad and E. M\"orstell, 
J. Cosmol. Astropart.Phys 09, (2004) 001. 
\bibitem{Corasaniti} P.S. Corasaniti et al., Phys. Rev. {\bf D70}, 0830006 (2004)
\bibitem{Basset} B.A. Basset, P.S. Corasaniti and M. Kunz, 
Astrophys.J. {\bf 617} L1-L4 (2004) 
\bibitem{LinderSNAP} see http://snap.lbl.gov or E. V. Linder, astro-ph/0406186
\bibitem{Ishak} M. Ishak, astro-ph/0501594


 
\end{thebibliography}
\end{document}